\documentclass[conference]{IEEEtran}
\IEEEoverridecommandlockouts
\usepackage{times}
\usepackage{amsmath,amssymb}
\usepackage{tikz}
\usetikzlibrary{arrows, shadows, positioning, patterns}
\usepackage{subcaption}
\usepackage{enumerate}
\usepackage{listings}
\usepackage{hyperref}
\usepackage[disable]{todonotes}
\usepackage{graphicx}
\usepackage{algorithm}
\usepackage{varwidth}
\usepackage{xspace}
\usepackage{url}
\usepackage[noend]{algpseudocode}

\usepackage{listings, xcolor}

\definecolor{verylightgray}{rgb}{.97,.97,.97}

\lstdefinelanguage{Solidity}{
	keywords=[1]{anonymous, assembly, assert, break, call, callcode, case, catch, class, constant, continue, contract, debugger, default, delegatecall, delete, do, else, event, export, external, false, finally, for, function, gas, if, implements, import, in, indexed, instanceof, interface, internal, is, length, library, log0, log1, log2, log3, log4, memory, modifier, new, payable, pragma, private, protected, public, pure, push, require, return, returns, revert, selfdestruct, send, storage, struct, suicide, super, switch, then, this, throw, true, try, typeof, using, value, view, while, with, addmod, ecrecover, keccak256, mulmod, ripemd160, sha256, sha3}, 
	keywordstyle=[1]\color{blue}\bfseries,
	keywords=[2]{address, bool, byte, bytes, bytes1, bytes2, bytes3, bytes4, bytes5, bytes6, bytes7, bytes8, bytes9, bytes10, bytes11, bytes12, bytes13, bytes14, bytes15, bytes16, bytes17, bytes18, bytes19, bytes20, bytes21, bytes22, bytes23, bytes24, bytes25, bytes26, bytes27, bytes28, bytes29, bytes30, bytes31, bytes32, enum, int, int8, int16, int24, int32, int40, int48, int56, int64, int72, int80, int88, int96, int104, int112, int120, int128, int136, int144, int152, int160, int168, int176, int184, int192, int200, int208, int216, int224, int232, int240, int248, int256, mapping, string, uint, uint8, uint16, uint24, uint32, uint40, uint48, uint56, uint64, uint72, uint80, uint88, uint96, uint104, uint112, uint120, uint128, uint136, uint144, uint152, uint160, uint168, uint176, uint184, uint192, uint200, uint208, uint216, uint224, uint232, uint240, uint248, uint256, var, void, ether, finney, szabo, wei, days, hours, minutes, seconds, weeks, years},	
	keywordstyle=[2]\color{teal}\bfseries,
	keywords=[3]{block, blockhash, coinbase, difficulty, gaslimit, number, timestamp, msg, data, gas, sender, sig, value, now, tx, gasprice, origin},	
	keywordstyle=[3]\color{violet}\bfseries,
	identifierstyle=\color{black},
	sensitive=false,
	comment=[l]{//},
	morecomment=[s]{/*}{*/},
	commentstyle=\color{gray}\ttfamily,
	stringstyle=\color{red}\ttfamily,
	morestring=[b]',
	morestring=[b]"
}

\usepackage[final]{changes}

\title{Smart Contracts on the Move\thanks{Preprint to appear in the 50th IEEE/IFIP Dependable Systems and Networks Conference (DSN'20)}}

\author{\IEEEauthorblockN{Enrique Fynn}
\IEEEauthorblockA{\textit{Universit\`{a} della Svizzera Italiana} \\
Switzerland}
\and
\IEEEauthorblockN{Alysson Bessani}
\IEEEauthorblockA{\textit{LASIGE, Faculdade de Ci\^{e}ncias} \\
\textit{Universidade de Lisboa} \\
Portugal}
\and
\IEEEauthorblockN{Fernando Pedone}
\IEEEauthorblockA{\textit{Universit\`{a} della Svizzera Italiana} \\
\textit{Interchain Foundation} \\
Switzerland}
}

\begin{document}
\maketitle
\begin{abstract}
   Blockchain systems have received much attention and promise to revolutionize many services. Yet, despite their popularity, current blockchain systems exist in isolation, that is, they cannot share information. While interoperability is crucial for blockchain to reach widespread adoption, it is difficult to achieve due to differences among existing blockchain technologies. This paper presents a technique to allow blockchain interoperability. The core idea is to provide a primitive operation to developers so that contracts and objects can switch from one blockchain to another, without breaking consistency and violating key blockchain properties. To validate our ideas, we implemented our protocol in two popular blockchain clients that use the Ethereum virtual machine. We discuss how to build applications using the proposed protocol and show examples of applications based on real use cases that can move across blockchains. To analyze the system performance we use a real trace from one of the most popular Ethereum applications and replay it in a multi-blockchain environment.
\end{abstract}

\section{Introduction}

Blockchain has gained much attention since the introduction of Bitcoin~\cite{bitcoin_nakamoto}, and
several blockchain systems have been developed thereafter~\cite{blockchain_in_the_wild}.
\added{Ideally} a blockchain is a geographically replicated state machine that tolerates
Byzantine failures.
Blocks in a blockchain contain transactions, usually cryptographically signed
by a user. 
In a permissionless blockchain system, blocks are produced by miners,
each block cryptographically linked to the previous
one, forming a chain. To produce a valid block, miners must solve a
cryptographic puzzle. The miner whose valid block makes it to the canonical
chain receives the block reward. 
In a permissioned blockchain system, miners can be in a consortium where they
behave similarly to members of traditional BFT algorithms, such as PBFT~\cite{pbft},
in which case they are named ``validators''.

As blockchain technology reaches mainstream use, it starts to face issues typical of more mature distributed systems technologies. 
Two formidable challenges are \emph{scalability} and \emph{interoperability}. 
Scalability has been early recognized as a major limitation of existing blockchain systems.
And several attempts have been made to improve blockchain performance (e.g.,~\cite{ouroboros, omniledger, rapidchain}).
In general, distributed systems scale performance by partitioning (sharding) the application state \cite{spanner}. 
If the partitioning is such that most application requests can be executed within a single partition, and the load among partitions is balanced, then performance scales with the number of partitions. 
Unfortunately, few applications can be optimally partitioned (i.e., all requests fall within a single partition and load is balanced among partitions).
As a result, most partitioned systems must handle requests that span multiple partitions. 
In the particular case of blockchain, it has been shown that even with a nearly perfect partitioning of the data, the existing Ethereum workload would result in a substantial number of cross-partition transactions~\cite{challenges_pitfalls}.
The Achilles heel of scalable blockchain systems is their ability to handle cross-partition transactions.

Interoperability has been in the blockchain wishlist for some time.
Yet, to date, no general mechanism has been proposed to share information across different blockchains.
Inter blockchain communication (IBC) is necessary for multiple blockchains to co-exist in a heterogeneous way.
\deleted{It is imperative for blockchains to reach global-scale adoption.}
For blockchains to interact with each other, some form of
synchronization is required across blockchains.
There are two main classes of solutions to handle transactions that involve multiple blockchains:
(a)~coordinating the blockchains involved in the execution of the transaction, in a scheme akin to atomic commitment~\cite{twophasecommit,hyperservice}; and
(b)~moving the state required by the transaction to a single target blockchain and then executing the transaction locally at the target blockchain.

Scalability and interoperability are different requirements.
However, they can be addressed with a common mechanism: 
a \emph{move operation} that allows accounts and arbitrary computation (i.e., \emph{smart contracts}) to consistently migrate from one blockchain to another.
In brief, the move operation works in two steps.
In the first step, it locks a smart contract in the source blockchain.
Once locked, the smart contract state cannot be changed in the source blockchain.
A second step recreates the smart contract in
the target blockchain in a provably correct way.
%

We have implemented
the Move protocol in Ethereum~\cite{ethereum_wp} and Burrow~\cite{burrow}, two popular blockchains, and evaluated it with different applications.
Even though smart contracts with arbitrary code can move across blockchains consistently,
we argue that developers should think of smart contracts as first-class objects that can move within blockchains.
For example, if a smart contract
maintains a set of users in its state, moving it will likely be
inefficient because a possibly large state with all users has to move
together with the smart contract. If, instead, the smart contract creates a
new smart contract per user, then the move can happen more efficiently, at the granularity of individual users.



%

\added{In summary,} the paper makes the following contributions:
\begin{itemize}
    \item We introduce the move operation, which allows programmable blockchains to interoperate, and provide a programming model for smart contract developers.
    \item We extend Solidity, a popular smart contract language, to include
    operations that help developers to program smart contracts that can move
    within blockchains. We propose an interface for token smart contracts and evaluate their usage.
    \item We modify two different blockchain clients and extensively analyze how the system behaves with different applications based
    on both synthetic and real workloads.
\end{itemize}

The rest of the paper is structured as follows.
Section~\ref{sec:background} presents the background necessary to understand
the move operation, detailed in Section~\ref{sec:move}.
Section~\ref{sec:applications} explains how to use the move operation
to implement interoperability and sharding.
In Section~\ref{sec:usecases}, we describe two different applications
that can benefit from the move operation, and from
Section~\ref{sec:deployment}~to~\ref{sec:ibc_exp} we report on the experimental
evaluation.
Finally, Section~\ref{sec:related_work} surveys previous work and in
Section~\ref{sec:conclusions} concludes the paper.

\section{Background} \label{sec:background}

A blockchain system is a distributed ledger, that is, an append-only log of
transactions. \emph{Clients} submit transactions to the blockchain, which are
appended to the log and then executed. Geographically distributed
\emph{nodes} interconnected through a peer-to-peer overlay network
implements the append-only log abstraction. Clients and nodes may be
\emph{honest}, in which case they follow their protocol specification, or
\emph{malicious}, in which case nothing can be assumed about their behavior.
The blockchain system behaves correctly as long as a fraction of the nodes,
typically more than two-thirds, are honest.

The append-only log is structured as a linked-list of blocks, each block
divided between a header and a body. The header contains, among other information,
a cryptographic link to the previous block. The body contains a list of
transactions, each transaction cryptographically signed by the client that
submitted it. The way the linked list of blocks is built leads to two
categories of blockchain systems. In the first category, there are blockchain systems (e.g., Bitcoin~\cite{bitcoin_nakamoto}, Ethereum~\cite{ethereum_wp})
that allow the chain of blocks to momentarily fork, that is, multiple blocks may
be linked to a block.
In the second category, there are blockchain systems that ensure
 a total order on linked blocks (e.g., Burrow/Tendermint~\cite{burrow}).
 These systems require nodes to agree on the next
block to be appended to the chain, and therefore solve consensus.
The techniques proposed in this paper apply to both categories.

Blockchain systems can also be distinguished by the nature of the operations they support.
Some blockchain systems limit transactions to distributed asset transfer,
while others allow transactions to perform
arbitrary computation (e.g., Ethereum, HyperLedger Fabric~\cite{hyperledger_fabric},
Cosmos/Tendermint~\cite{Buchman2018}). In this paper, we assume blockchain systems in the second
category. 
In our model, we make a distinction between two types of data objects: accounts which hold state and have
a cryptographically derived unique identifier, and smart contracts, or for
short contracts.
Smart contracts encapsulate executable code, which can hold, read, and modify
their own state and call other smart contracts. Clients hold one asymmetric
key-pair per account, and every transaction originates from a client, who
provides proof of ownership of an account.

%
%


Blockchains can grow large in size and complexity.
For instance, Ethereum has
over three terabytes of log at the moment and it can take several weeks to
re-execute all appended transactions.
Clients with low storage or computational power
\deleted{are still able to verify}
\added{can succinctly prove the validity of an arbitrary piece of the state \cite{merkle_tree}}
without re-executing the entire log,
\added{provided they maintain}
\deleted{Provided that clients maintain}
and verify all the block headers.
\deleted{They can succinctly prove the validity of an arbitrary piece of the state \cite{merkle_tree}.}

Blockchain systems typically rely on a Merkle-tree or similar data structure to provide data integrity
checks.
For example, Bitcoin uses a binary Merkle-tree~\cite{merkle_tree}, while
Tendermint uses a modified AVL tree~\cite{iavlplus}. For the sake of
simplicity, we call these structures ``Merkle-trees''.
Each block header includes the root of a Merkle-tree (i.e., Merkle-root).
Data is encoded on the leaves of the Merkle-tree, and parent nodes are labeled
with the cryptographic hash of child nodes grouped together, compacting the
structure until a unique Merkle-root is reached. The objective of such data
structure is to provide a computationally and spatially cheap way to prove the integrity of
leaves of the state without necessarily having all the state.

Merkle-trees allow for a peer to hold only block headers and forego
downloading all the blockchain state.
Peers can ask for a proved piece of
partial state (Merkle-proof) at a specific block height from peers that have the state at the requested
block height or happen to have the same Merkle-proof.
The information provided by
these peers can be checked with the Merkle-root stored in the trusted block
header.
We denote the unique path of a valid Merkle-proof from object $v$ to
Merkle-root $m$ as $\{ v \} \mapsto m$. The leafs $v$ and nodes $h \in (\{ v \} \mapsto m)$ needed
to reconstruct the proof must be given to verify the
validity of the proof. 
The verification of Merkle-proofs can be done optimally in logarithmic time and space
on the number of nodes of the tree~\cite{merkle_tree}.

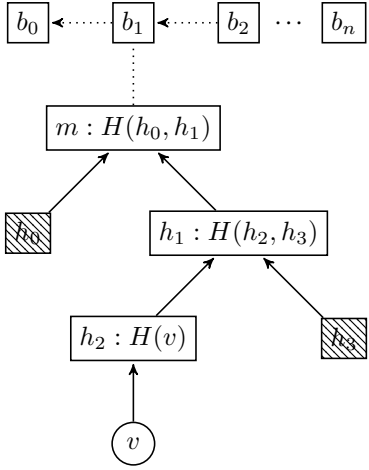
\begin{figure}[h]
  \centering
  \begin{tikzpicture}[->,>=stealth',shorten >=1pt,auto,node distance=1.4cm,
    semithick]
    \node[draw=black] (B0) at (0,0) {$b_0$};
    \node[draw=black, right of=B0] (B1) {$b_1$};
    \node[draw=black, right of=B1] (B2) {$b_2$};
    \node[draw=black, right of=B2] (B3) {$b_n$};

    \path [->, dotted](B1) edge node[] {} (B0);
    \path [->, dotted](B2) edge node[] {} (B1);
    \path (B2) -- node[auto=false]{\ldots} (B3);

    \node[draw=black, below of=B1] (C) {$m : H(h_0, h_1)$};
    \path [-, dotted](C) edge node[right] {} (B1);
    
    \node[draw=black, below of=C, right of=C] (H1) {$h_1 : H(h_2, h_3)$};
    \node[pattern=north west lines, draw=black, below of=C, left of=C] (H0) {$h_0$};
    \path [->](H0) edge node[] {} (C);
    \path [->](H1) edge node[] {} (C);

    \node[pattern=north west lines, draw=black, below of=H1, right of=H1] (H2) {$h_3$};
    \node[draw=black, below of=H1, left of=H1] (H3) {$h_2 : H(v)$};
    \path [->](H2) edge node[] {} (H1);
    \path [->](H3) edge node[] {} (H1);

    \node[shape=circle, draw=black, below of=H3] (V) {$v$}; 
    \path [->](V) edge node[] {} (H3);
  \end{tikzpicture}
  \caption{Merkle-proof example, $h_0$, $h_3$ $\in (\{v\} \mapsto m)$.}
  \label{fig:merkle_proof}
\end{figure}

Figure~\ref{fig:merkle_proof} illustrates how Merkle-proofs can
prune parts of the tree logarithmically.
Blocks from $b_0$ to $b_n$ are shown linked together
in the top. From block $b_1$ we see the Merkle-proof for $\{v\} \mapsto m$.
Given a hash function $H$, $m$ is $b_1$'s Merkle-root composed by $h_0$ and $h_1$
hashes. In the figure we see the result of asking for the Merkle-proof of $v$,
anyone can compute the Merkle-root of this proof and accept it only if it is equal to the trusted Merkle-root $m$ stored in $b_1$.
Observe that only $v$ and hatched nodes $h_0$ and $h_3$ are needed to \replaced{verify}{reconstruct} $m$.

%

\section{The Move protocol} \label{sec:move}
We propose a new method to move a contract from one blockchain to another.
In the following, we state the assumptions needed to support the move operation (Section \ref{sec:assumptions}),
introduce the general idea (Section \ref{sec:protocol}), present the move operation in detail (Section \ref{sec:pmodel}), and discuss extensions to the basic protocol (Section \ref{ref:currency}).

%
\subsection{Assumptions}
\label{sec:assumptions}

We make the assumptions listed next in order for blockchains to support the move operation.
Although these assumptions are not strictly necessary to move objects across blockchains, they simplify the Move protocol.

In particular, we assume that blockchains must:
\begin{enumerate}[(a)]
  \item support smart contracts (i.e., arbitrary computation);
  \item use the same execution environment (i.e., virtual machine); and
  \item provide a succinct way to prove state variables (e.g., using a Merkle-tree).
\end{enumerate}



Supporting smart contracts that can execute arbitrary computation allows us
to investigate more complex and generic use cases for the protocol.
When moved to the target blockchain, we assume that smart contracts can
execute the same opcode instructions they executed in the source blockchain,
i.e., they use the same execution environment.
This simplifies how communicating blockchains are capable
to understand each other.

Clients can have information about the Merkle-root of any other blockchain by downloading the correspondent block header.
Clients can listen to headers from multiple blockchains all at once. Block
headers have a constant size of usually hundreds of bytes and are on average a small fraction of block bodies.
For example, in Ethereum block headers are around 2\% of the block's body.
Moreover, blockchains willing to support the Move protocol must agree on certain
configured parameters discussed in Section~\ref{sec:ibc}.


\subsection{Overview}
\label{sec:protocol}

The state of a contract is presumably indivisible and must reside as a whole in a blockchain.
Accounts and smart contracts are restrained to live in a single blockchain at a time.
Therefore, we must ensure that if a contract moves from one blockchain (source) to another (target), it will no longer be ``active" in the source blockchain.
When the contract becomes active in the target blockchain, its state must be identical to its state when it became inactive in the source blockchain.
This implies that the move operation involving two blockchains must be atomic.

One way to implement an atomic Move operation is to resort to the well-known two-phase commit (2PC) protocol~\cite{twophasecommit}, or one of its more resilient variations (e.g., \cite{GL05}).
We refrain from using a 2PC-like Move protocol since it would introduce expensive coordination between the involved blockchains (e.g., members of each blockchain would have to exchange votes in a reliable manner).
Instead, we use a two-step approach that divides the move operation into two transactions, \emph{Move1} and \emph{Move2}. 
\deleted{The target blockchain the smart contract is moving to should always be specified in Move1, otherwise Move1 and Move2 can take an arbitrarily number of arguments.}+
In Move1, the state of a smart contract
is ``locked'' in the source blockchain, after which it is guaranteed not to be changed---although transactions can still read the contents of the locked smart contract.
In Move2, the smart contract is reconstructed in the target blockchain, after which it can be safely used.
To avoid simple attack vectors, the Move2 transaction is only successful at the target blockchain if it contains a proof that the Move1 transaction was successfully executed at the source blockchain.

This two-step approach reduces coordination between the source and the destination blockchains, but it complicates atomicity.
For example, the move of a contract can remain unfinished if the client fails after submitting the Move1 transaction and before it submits the corresponding Move2 transaction.
To account for such cases, we allow any client to execute the Move2 transaction, and thereby complete a possibly unfinished Move operation.
In the normal case, however, we expect the same client to execute both transactions.


\subsection{The Move protocol in detail}
\label{sec:pmodel}

Algorithm~\ref{alg:move} details the move operation of contract $c$ from blockchain $B_i$ to $B_j$.
We add a new field to a contract state, referred to as $L_c$, to
$L_c$ represent the blockchain identifier the smart contract $c$ currently resides in.
At low level, assigning a new value to $L_c$ in Move1 is implemented with a new EVM opcode, OP\_MOVE. 
The OP\_MOVE opcode takes as
argument the target blockchain identifier the smart contract is moving to (in this case $B_j$). When
OP\_MOVE is executed in $c$, it changes $L_c$ to $B_j$, and by consequence
blocks the contract state at $B_i$.
Any transactions that try to alter the state of blocked contract $c$ in $B_i$ will be aborted.

Move2 assumes the existence of two boolean functions,
$V_S$ and $V_P$.
$V_S(B, m)$ returns true if $m$ is a \added{valid} Merkle-root \deleted{accepted} in the blockchain $B$.
$V_P(V \mapsto m)$ returns true if the state $V$ of $c$ is proved by $V \mapsto m$.
The smart contract code and other blockchain specific variables (e.g., the amount
of currency held by the smart contract) are omitted in the
algorithm but still need to be proved by $V \mapsto m$.
Notice that before submitting a Move2 transaction for contract $c$, the client must acquire the proof $V \mapsto m$ at the source blockchain (discussed later).

The Move1 and Move2 transactions allow application developers to execute special routines when a contract is moved.
We illustrate the use of this functionality in the next section.

  \begin{algorithm}
  \footnotesize
    \caption{The operations.}\label{alg:move}
    \begin{algorithmic}[1]
    \Procedure{Move1}{$c, B_j$} \Comment{Move $c$ in $B_i$ to $B_j$, executed at $B_i$}
    \State $moveTo(\cdot)$ \Comment{Execute custom function}
    \State $L_c \gets B_j$ \Comment{Block contract $c$ in $B_i$}
    \EndProcedure
  
    \Procedure{Move2}{$c, V \mapsto m$} \Comment{Complete move of $c$, execute at $B_j$}
  
  
    \If{$L_c \neq B_j$} \Comment{Is $c$ being moved to the wrong blockchain?}
      \State \Return abort
    \EndIf
    
    \If{$V_S(B_i, m) = \mathit{false}$} \Comment{Invalid Merkle-root}
      \State \Return abort
    \EndIf

    \If{$V_P(V \mapsto m) = \mathit{false}$} \Comment{Invalid proof}
      \State \Return abort
    \EndIf
  
    \For{all $v \in V$}
      \State Call SSTORE($v$.key $v$.value) \Comment{Recreate storage in $B_j$}
    \EndFor
  
    \State \Return $moveFinish(\cdot)$ \Comment{Execute custom function}
  
    \EndProcedure
    \end{algorithmic}
  \end{algorithm}

\subsection{A concrete implementation}


We have \replaced{integrated}{introduced} the Move operation in the Solidity programming language~\cite{solidity}.
In our prototype, smart contract developers must implement two
functions to allow contracts to move, $moveTo(\cdot)$ and
$moveFinish(\cdot)$ (see Algorithm~\ref{alg:move}). 
This provides the application developer a great deal of flexibility.
For example, in Listing~\ref{code:moveTo} we have an excerpt of Solidity code that in few lines ensures that only the contract's owner is allowed to move the
contract and the contract must remain at least three days in the target
blockchain before moved again.

\begin{figure}[t]
  \begin{lstlisting}[caption={Excerpt of a movable contract in Solidity.},language=Solidity, label={code:moveTo}]
  address owner;
  uint movedAt;
    function moveTo(uint _blockchainId) public {
      require(owner == msg.sender);
      require(now - movedAt >= 3 days);
    }
    function moveFinish() public {
      movedAt = now;
    }
  \end{lstlisting}
\end{figure}

\subsection{Preventing replay attacks}

If a client executes transaction $T_{move1}$ at blockchain $B_i$ to move contract $c$ to $B_j$,
any client can craft a special transaction $T_{move2}$ to be executed in blockchain $B_j$
that reconstructs the state of $c$ in $B_j$. In $T_{move2}$ the client
appends the state of contract $c$ encoded as $V$ in the Merkle-proof $V \mapsto m$.
Target blockchain $B_j$ is responsible for verifying that $m$ is accepted by the blockchain as
a valid Merkle-root of $B_i$.
Nodes in blockchain $B_j$ verify the correctness of $c$'s state by verifying
$V \mapsto m$ and $B_i$'s state root hash. If the proofs are valid, $c$'s state
can be safely reconstructed in $B_j$.

\begin{figure}[t]
  \centering
  \includegraphics[width=\linewidth]{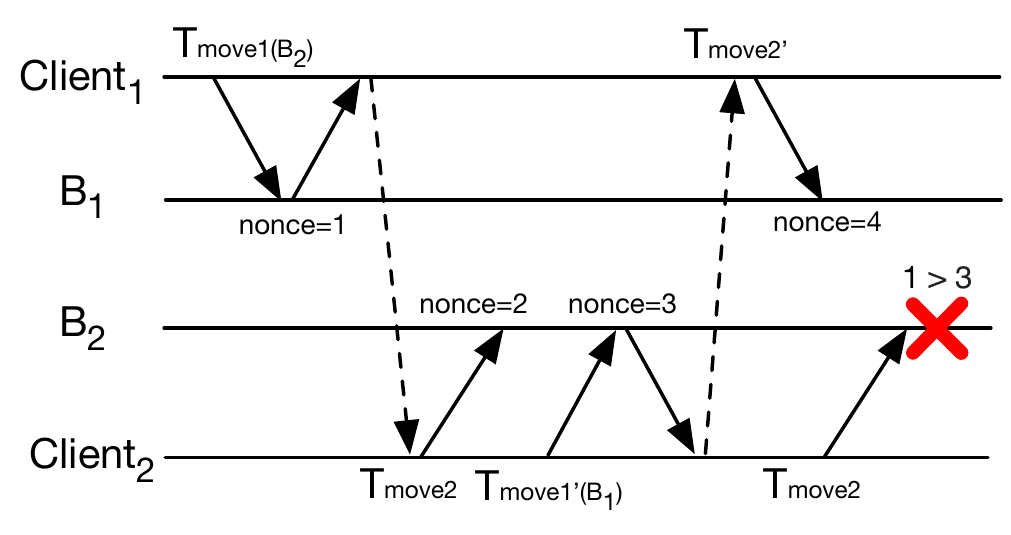}
  \caption{Preventing replay on stale data.}
  \label{fig:stale_data_prevention}
\end{figure}

Additional measures should be taken to prevent replay attacks.
The attack consists in a (malicious) client crafting a $T_{move2}$ transaction that uses old state of contract $c$.
The replayed transaction would obviously lead to inconsistencies as transactions that followed the first (and thus legitimate) Move2 transaction would be lost.
One remedy would be to have nodes store the contract's nonce, a monotonically increasing number that is increased every time
the contract is invoked.
For instance, in Figure~\ref{fig:stale_data_prevention} a contract is moved
from $B_1$ to $B_2$ (transactions $T_{move1}$ and $T_{move2}$, respectively) and afterwards back to $B_1$ (transactions $T_{move1'}$ and $T_{move2'}$).
It starts with nonce (n) equal to zero and as soon
as $T_{move2}$ is executed in $B_2$ it increments
the nonce by one.
Afterwards $T_{move1'}(B_1)$ completes in $B_2$ and changes the nonce to three.
When client$_2$ tries to replay transaction $T_{move2}$, the
contract's nonce is one which is less than what was previously seen by $B_2$ and
the transaction aborts.

\subsection{Handling currencies} \label{sec:atomic_swap}
\label{ref:currency}

Most cryptocurrencies rely on an internal currency for generating incentives
to maintain the system (e.g., paying miners and transaction fees).
In some cases, the currency is the main usage of the system (e.g., Bitcoin).
It is important therefore to have a way of transferring this part of the system state from one
blockchain to another.
As it turns out, we can devise a simple mechanism that
uses our protocol to accomplish the feat. It suffices for
smart contracts to be allowed to hold currency in their state.
We can then implement smart contract ``relays'' that can transfer currency from blockchains
by creating a token in the target blockchain that is provably locked in the
source blockchain, similar to Pegged~Side~Chains~\cite{pegged_sidechains}.
Currencies can be unlocked when contracts return to the original blockchain.

Assume for example that we would like to transfer $e$
units of currency from $client_1$ to $client_2$, from blockchain $B_i$
to blockchain $B_j$.
We assume the existence of contract $c$ in $B_i$,
that when called by client $client_1$ with input $B_j$, $client_2$ and value associated
$e$ creates a contract $r$ that has $e$ units of currency and
lets $client_2$ withdraw $e$ from its state (i.e., it executes Move1($B_j$) on creation).
The $client_2$ can call $T_{move2}$ on
$r$ effectively moving it to $B_j$ where the funds would be available
as a $B_i$'s token in $B_j$.

\begin{figure}[t]
  \centering
  \includegraphics[width=\linewidth]{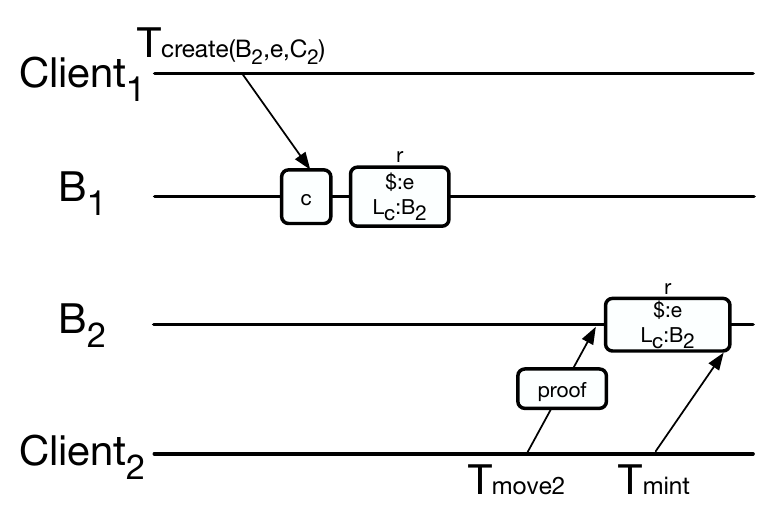}
  \caption{Move operation example.}
  \label{fig:move_ex}
\end{figure}

In Figure~\ref{fig:move_ex}, we can see an example where a client successfully transfers
a currency token from blockchain $B_1$ to a token representation in blockchain $B_2$.
Contract $c$'s function $create$ is called with $e$ units of $B_1$'s associated currency with $T_{create}$.
The transaction creates contract $r$ with $e$ units of currency, seen in the figure as ``\$'', afterwards
the same function calls $r$'s $moveTo(B_2)$ changing $r$'s $L_c$ to $B_2$.
The newly created contract has functions to generate tokens which are proved to be backed by $e$
in $B_1$.
The $client_2$ waits for the transaction inclusion in $B_1$
and sends transaction $T_{move2}$ to $B_2$, proving that $r$ was moved to $B_2$.
After transaction $T_{move2}$ is included in $B_2$,
$client_2$ calls transaction $T_{mint}$ which executes code in $r$ creating tokens in $B_2$ that represent
the locked coins in $B_1$.

\subsection{Additional details}
\label{ref:adetails}

\paragraph{Account identifiers}

Although each blockchain maintains its own set of accounts, its identifier
can be the same if the interacting blockchains use the same rule to derive
such identifiers. Consequently, clients could use the same cryptographic keys
to use accounts in different blockchains. It becomes essential to
incorporate the blockchain's identification to functions that compute
contract addresses to ensure a unique system-wide contract identification to avoid
collisions in contract identifiers.

\paragraph{Finding contracts}
$L_c$ can only have two logical states: either in the current blockchain or
transferred (executed Move1), and if $moveTo(\cdot)$ and $moveFinish(\cdot)$ are
implemented correctly, it can always go back to the first state with Move2.
A client who does not know where contract $c$ is located can use $L_c$ to
track the contract's location every time it moves.

\paragraph{Stale data}
Every time a contract is moved it leaves behind stale state on the original blockchain,
which could \replaced{be garbage collected}{get removed}, paying attention to guard against the attack
previously described.
Designing fee incentives to clean the state is left as future work.

\begin{figure*}[t!]
  \begin{lstlisting}[caption={Scalable Token interfaces extending ERC20.},language=Solidity, label={code:stoken}]
  contract STokenI {
    function totalSupply() public view returns (uint);
    function newAccount() public payable returns (AccountI, uint);
    function newAccountFor(address _forAddr) public payable returns (AccountI, uint);
    event CreatedAccount(address account, uint salt);
  }
  
  contract AccountI {
    function balance() public view returns (uint);
    function allowance(address _spender) public view returns (uint);
    function transfer(AccountInterface _to, uint _tokens) public returns (bool);
    function approve(address _spender, uint _tokens) public returns (bool);
    function transferFrom(AccountInterface _to, uint _tokens) public returns (bool);
    function debit(uint _tokens, bytes _proof) public returns (bool);
    function moveTo(uint _shardId) public;
    function moveFinish() public;
    event Transfer(address _to, uint _tokens);
    event Approval(address _spender, uint _tokens);
  }
  \end{lstlisting}
\end{figure*}
\section{Applications} \label{sec:applications}
The Move protocol can be used by blockchains to implement two important
blockchain concepts: interoperability and sharding.
Interoperability and sharding are related concepts that have increased in importance as blockchain technology matures.



\subsection{Interoperability} \label{sec:ibc}

Blockchains coexist nowadays with different transaction designs, cryptographic features,
and trust assumptions in a heterogeneous environment.
Interoperability provides a way to offload transactions from one blockchain to another,
unleashing the potential to scale different applications and experiment with different
combinations of blockchains.

Interoperability in permissionless systems is challenging mainly because forks can occur since
block propagation time is unbounded~\cite{propagation_btc}, which
invalidates transactions that build on the losing side of the fork. A way for
systems proposing interoperability~\cite{pegged_sidechains, polkadot, cosmos}
to interact with permissionless blockchains is by introducing a parameter $p$
that specifies the minimum number of blocks that a transaction's block should be behind the blockchain's
head for it to be accepted by the other blockchain.
The parameter can be configured according to each blockchain involved in
the interoperability protocol.

Miners or validators of blockchains willing to interoperate should maintain a
light client that validates merkle-roots of other blockchains, proposals for this
scheme are discussed in Section~\ref{sec:related_work}.
\deleted{We can implement interoperability by using the Move protocol, provided
assumptions discussed above are met.}


\subsection{Sharding}
Sharding enables the blockchain state to be divided into shards responsible
for holding a certain portion of the state. Sharding 
preserves some of the blockchain assumptions, but there are clear trade-offs
in security because the members themselves have to be sharded.
The way objects are assigned to each shard plays an important role when
sharding a blockchain for scalability. For instance, if objects are  
randomly partitioned into shards, most of the transactions will likely be cross-shard.
The rate of cross-shard transactions also increases with the number of
shards, and sharding the state while minimizing the number of cross-shard
transactions keeping the various shards balanced is a hard
problem~\cite{challenges_pitfalls}.

To cope with changes in load it becomes essential to incorporate a method to
move state from shards, offloading one shard in detriment of another.
As shards get congested and fees increase, users are tempted to move their
contracts to underused shards.

\section{Use cases} \label{sec:usecases}
\newcommand{\cryptokitties}{\textit{ScalableKitties}\xspace}
\newcommand{\ethprice}{144\xspace}

We implemented two applications for smart contracts and show how they scale with
the number of blockchains:
\begin{itemize}
  \item \textit{SCoin}: A token smart contract based on a popular Ethereum token interface.
  \item \cryptokitties: A clone of \textit{CryptoKitties}, a popular Ethereum application where virtual cats can migrate and reproduce in different shards.
\end{itemize}

\subsection{SCoin} \label{sec:scalable_coin}

ERC20~\cite{erc20} is a standard interface widely used in
Ethereum for token operations including token transfers.
\textit{STokenI} and \textit{AccountI}, defined in Listing~\ref{code:stoken}, are interfaces that support all ERC20 operations
and allow for contracts to move from one blockchain to another.
The main idea of the interface is to use one instance of \textit{AccountI} per user account.
Typical ERC20 implementations hold token balances in a map data
structure, which multiple blockchains cannot share in our design since
we do not allow for contracts to live in two or more blockchains at the same time.

Once created, accounts can freely move
from blockchain to blockchain using the \emph{moveTo}  and \emph{moveFinish} functions.
It is left to the developer to restrict or even define a policy for moving accounts
between blockchains.

To illustrate the \textit{STokenI} interface, we implement \textit{SCoin}, a scalable token contract
that implements \textit{STokenI}.
\added{\textit{SCoin} creates instances of \textit{SAccount}, which implements \textit{SAccountI}}.
The implementation of \textit{SCoin} and most functions of \textit{SAccount} are
straightforward and application-dependent.
We focus next on how to do safe
transfers between one \textit{SAccount} to another.
To execute a transaction that transfers $e$ tokens from \textit{SAccount} $A$ to $B$,
contract $A$ has to decrease $e$ from its state (called by \textit{transfer} function)
and $B$ has to increase $e$.
This is done by \deleted{having the same transaction call}\added{calling} the \textit{debit}
function in $B$.
If contracts $A$ and $B$ are in different blockchains, they have
to be first moved to the same blockchain to be able to call each other.
Once both contracts are in the same blockchain, how can $A$ know that $B$ is
what it claims to be? For instance, one could design a contract $B$ that,
when \textit{debit} is called, increases the contract's tokens by an arbitrary amount.
$A$ could ask its parent if $B$ was created by the same contract, but
$A$'s parent contract might be in a different blockchain.
The interface does not specify how contract $A$ can be sure of contract $B$'s origin.
It is up to the developer to devise safeguards to prevent incorrect usage.
The key idea for \textit{SCoin} is holding a proof in $B$ that it was created
by the same contract that created $A$.

When we create an instance of \textit{SAccount} in \textit{SCoin} it uses a monotonically increasing salt,
stored in the instance state. The salt is used to calculate the identifiers of both contracts using the
\emph{create2} opcode~\cite{ethereum_gavin}.
With $A$'s salt, $B$ can attest that $A$ was created by the same contract that created $B$ and vice-versa.
To execute a transfer from \textit{SAccount} $A$ to $B$, contract $A$ attests $B$'s origin,
decrements its own balance and calls the function $debit(\cdot)$ in
$B$. Contract $B$ agrees to debit its own account only if $A$ passes the same check,
\replaced{and}{if the check passes,} $A$ can safely add $B$'s fund to its own balance.
The checks of origin are done with one inexpensive hash operation.
In our implementation case, we take advantage of the way contract identifiers are generated in the EVM.
A more generic method could be devised using Merkle proofs with the same proposed interfaces.


\subsection{\cryptokitties} \label{sec:cryptokitties}
\todo[inline]{How to cite +4m tx in CK?}
\cryptokitties is a \replaced{clone}{copy} of \textit{CryptoKitties}, a popular Ethereum smart contract
that was created in November 23th 2017 and until the writing of this paper
had over four million \replaced{related transactions}{transactions on it}. During the apex of its popularity,
\textit{CryptoKitties} \deleted{was extensively used and} congested the Ethereum network for several days,
accounting for over fifteen percent of all Ethereum transactions~\cite{cryptokitties}.
%
In \textit{CryptoKitties} cats are collectibles that can be bred to generate more cats
following a set of rules (e.g., sibling cats cannot mate). Cats were first
generated by the contract's owner. Both the initial and bred cats are
sold in an auction smart contract. The game was the first mainstream application
built on top of Ethereum, and some cats were sold for more than a hundred thousand
dollars at the time.\replaced{
The contract is still actively used today.}{and the contract is still actively used until the writing of this paper.}

\cryptokitties's functions map one-to-one to the \textit{CryptoKitties} smart
contract but for simplicity we discuss only the functions that are related to the execution of cross-blockchain transactions. Cats are created in
two ways: either by having the contract's owner calling a function to
generate ``promotional'' cats or by breeding two cats to generate a third.
Breeding is the only operation that can generate cross-blockchain transactions
because bred cats can be in different blockchains and need to be moved to
the same one. Furthermore, if the owner of cat $A$ wants to breed $A$ with
$B$, it either needs to own both cats or $B$'s owner has to permit $B$ to be sired
with $A$. In the experiments in Section~\ref{sec:exp_cryptokitties} we
replay transactions from the \textit{CryptoKitties} contract on the \cryptokitties contract.

\section{Deployment} \label{sec:deployment}

We modify Hyperledger Burrow~\cite{burrow} and Ethereum~\cite{go_ethereum} to implement the protocol defined
in Section~\ref{sec:move}. The resulting systems allow blockchains to communicate with each other (IBC) or implement sharding.
To validate our approach, we conducted two types of experiments: we shard Burrow and analyze
how applications can scale performance, and make smart contracts migrate from Ethereum to Burrow and
vice-versa to assess the performance and monetary costs of IBC.
Both Burrow and Ethereum implement the EVM model, where each opcode executed by the smart contracts has a cost modeled in \emph{gas}, e.g., a sum between two integers costs 3 gas, while creating a new smart contract costs 32000 gas~\cite{ethereum_gavin}.

Burrow uses Tendermint for consensus, which by design
introduces the application's Merkle-root from block $n$ in block
$n+1$, but in Burrow the state of block $n$ is saved only in block $n+1$, therefore there
is a need to wait for two blocks to prove the transaction inclusion required
by Move2.
For the experiments we set $p$ (defined in Section~\ref{sec:ibc}) equal to two
blocks in Burrow, since clients have no option other to wait for two blocks to
get the proof of inclusion of a Move1 transaction.
For Ethereum we set $p$ to six blocks.
\deleted{, as recommended in Nakamoto consensus systems to ensure transaction finality with high probability.}

Tendermint is configured to wait for five seconds between each consecutive block, the
observed latency being slightly higher than this.
Ethereum is configured to wait 15 seconds, which is similar to Ethereum's main public network.

All experiments were conducted using a heterogeneous cluster in a local area
network with simulated latencies between nodes based on the values published in \cite{redbelly}, 
where the authors evaluate nodes in 14 regions in four
continents on Amazon data centers. We emulate this environment in the cluster
and randomly allocate nodes to regions.

\section{Sharding Experiments} \label{sec:sharding_exp}

We evaluate the Move protocol using the two applications described in
Section~\ref{sec:usecases}.
The objective is to show how the capacity to move smart contracts
can significantly improve the performance of the applications.
In all sharding experiments, one node hosts all clients and maintains one connection
per shard to broadcast the client's transactions.
We decided to run one validator in each node and 10 validators per shard due
to a limitation of our cluster size, comprised of 80 computers\added{, each one with an eight-core Intel Xeon L5420 processor working at 2.5GHz, 8Gb of memory, SATA SSD disks and 1Gbps ethernet card.}
With this configuration we can run a maximum of 8 shards and we decided to run
experiments with 1 (no sharding), 2, 4 and 8 shards.
To decide which shard to send the contracts to when needed, we use
hash partitioning where the contract's shard is decided by the hash of the contract's
identification. Using hash partitioning ensures a good balance among shards
but implies probably a higher cross-shard rate the more shards there are.
We do not focus our attention to examine different partitioning techniques but we believe
greater improvements are possible by using different sharding methods~\cite{challenges_pitfalls}.

\subsection{\cryptokitties} \label{sec:exp_cryptokitties}

In order to produce the data for experiments \deleted{in Section~\ref{sec:exp_cryptokitties}}
we scanned all transactions involving the \textit{CryptoKitties} contract
deployed~\footnote{At address 0x06012c8cf97bead5deae237070f9587f8e7a266d} in
Ethereum since its inception.

\begin{figure}
  \centering
  \begin{tikzpicture}[->,>=stealth',shorten >=1pt,auto,node distance=1.4cm,
    semithick]
    \node[shape=circle,draw=black] (C1) at (0,0) {Tx$_1$};
    \node[shape=circle,draw=black, right of=C1] (C2) {Tx$_2$};
    \node[shape=circle,draw=black, below of=C2] (AC2) {Tx$_3$};
    \node[shape=circle,draw=black, below of=C1] (BC1C2) {Tx$_4$};
    \node[draw=black, above of=C1] (I1) {$c_1$};
    \node[draw=black, above of=C2] (I2) {$c_2$};
    \node[draw=black, left of=I1] (I3) {$c_3$};

    \path [->](BC1C2) edge node[right] {} (AC2);
    \path [->](BC1C2) edge node[right] {} (C1);
    \path [->](AC2) edge node[right] {} (C2);

    \path [->](I1) edge node[right] {} (C1);
    \path [->](I2) edge node[right] {} (C2);
    \path [->](I3) edge node[right] {} (BC1C2);
  \end{tikzpicture}
  \caption{Dependency graph example}
  \label{fig:dependency_graph}
\end{figure}
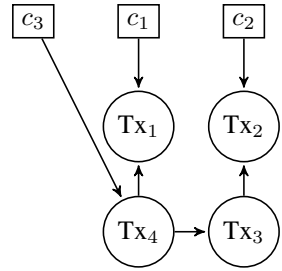

Since every transaction from the original contract must succeed in our implementation,
we first construct a dependency DAG \added{(Directed Acyclic Graph)} of all the transactions and execute
them respecting their dependencies.
By doing so we are able to execute some of the transactions in parallel.
for instance, consider a client that owns cat $c_1$ and wants
to breed its cat with another user's cat $c_2$. To execute this transaction, the transaction
graph of Figure~\ref{fig:dependency_graph} needs to be respected.
First cats $c_1$ and $c_2$ have to be created with Tx$_1$ and Tx$_2$, respectively, then $c_2$'s owner agrees with
the breeding with Tx$_3$ and finally Tx$_4$ breeds $c_1$ and $c_2$, creating $c_3$.
Vertices $c_1$, $c_2$, and $c_3$ are pointers to transaction vertices that have a dependency
on $c_1$, $c_2$, and $c_3$, respectively.
Notice that leaf transactions in the DAG can be executed in parallel, for instance,
Tx$_1$ and Tx$_2$ can be executed in parallel but Tx$_4$ can only
execute when it becomes a leaf, i.e., both Tx$_1$ and Tx$_3$ finish.

We replay \textit{CryptoKitties} transactions on
\cryptokitties in a sharded environment running multiple
instances of the Burrow client.
We use the dependency DAG described previously to replay transactions to the contract.
Transactions that are appended in a block (executed) are removed from the DAG. Subsequent transactions
that become leaves in the DAG are submitted until a limit of 250 outstanding transactions is reached.
We pre-process the whole DAG in memory, broadcasting the first transactions,
updating the DAG, and possibly sending other transactions whose dependencies are satisfied.
The process continues until the experiment is over.
Increasing the number of shards leads to an increase in the number of cross-shard
transactions, that in turn reduce the number of leaves in the DAG since cross-shard
transactions depend at least on two transactions: Move1 and Move2.

\begin{figure*}[htb]
  \centering
  \includegraphics[width=0.48\linewidth]{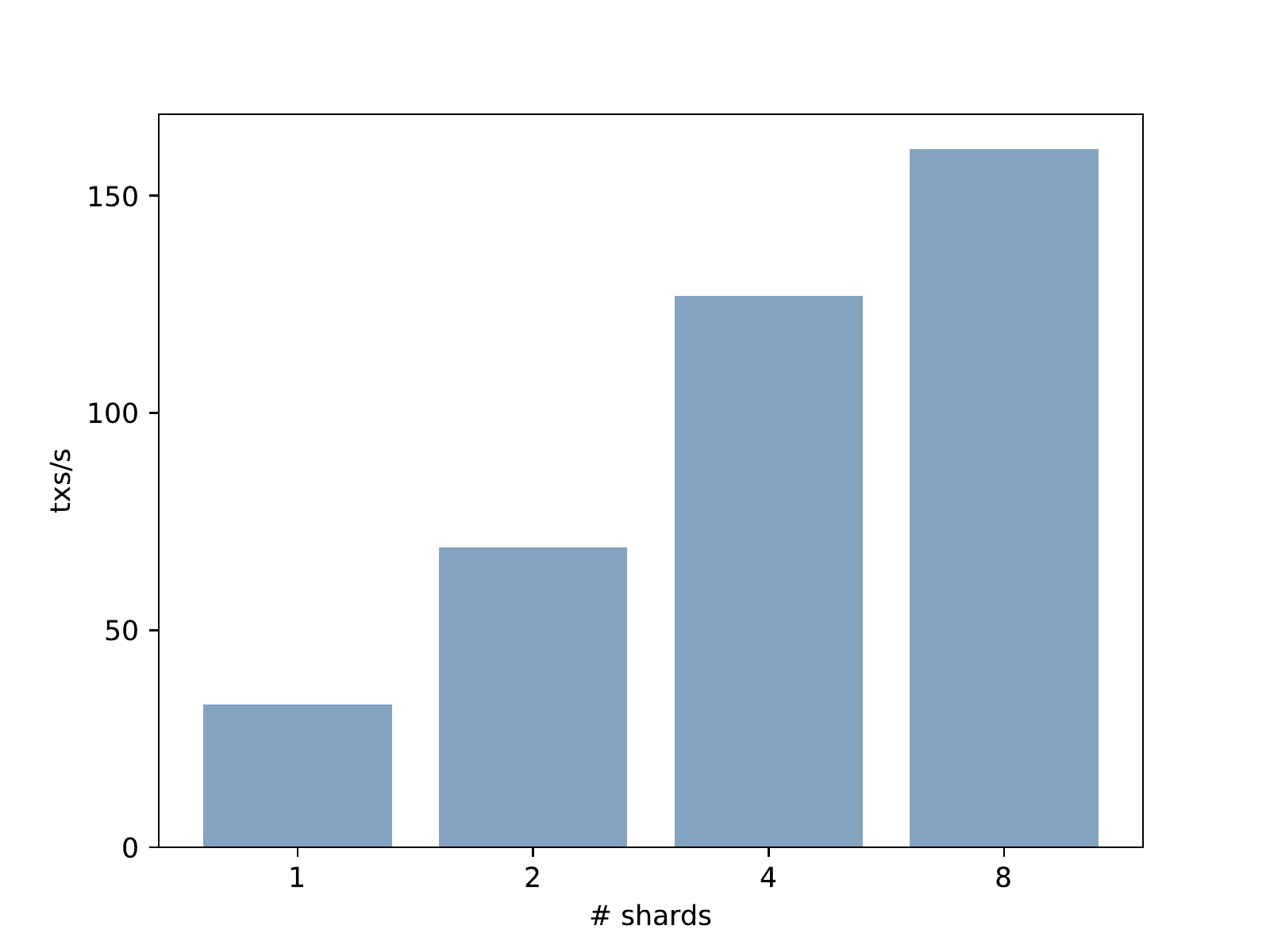}
  \includegraphics[width=0.48\linewidth]{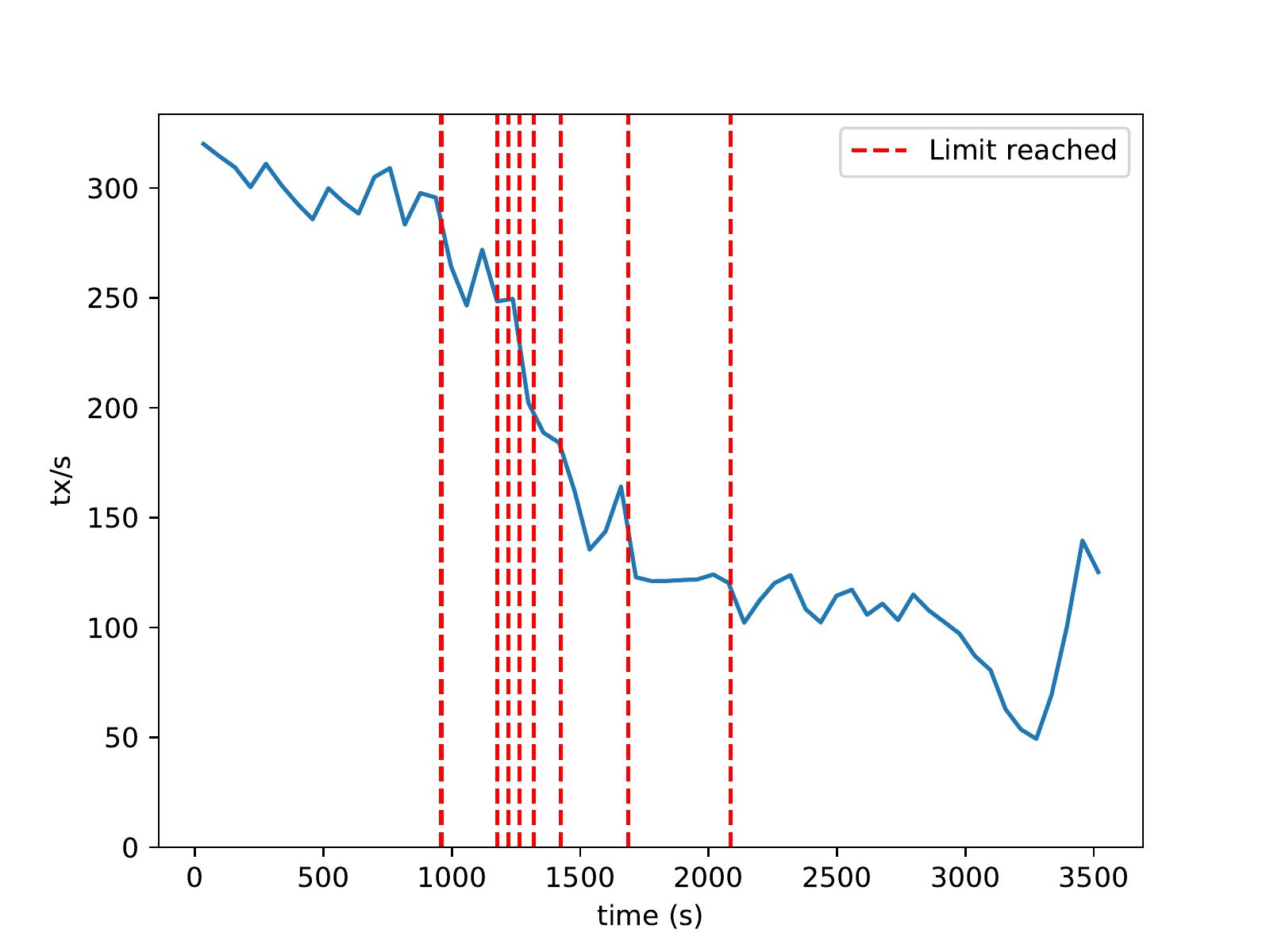}
  \caption{\cryptokitties throughput for 2, 4 and 8 shards (left), and aggregated throughput over time for 8 shards (right).
  }
  
  \label{fig:ck_tput}
\end{figure*}

\begin{figure*}[htb]
  \centering
  \includegraphics[width=\linewidth]{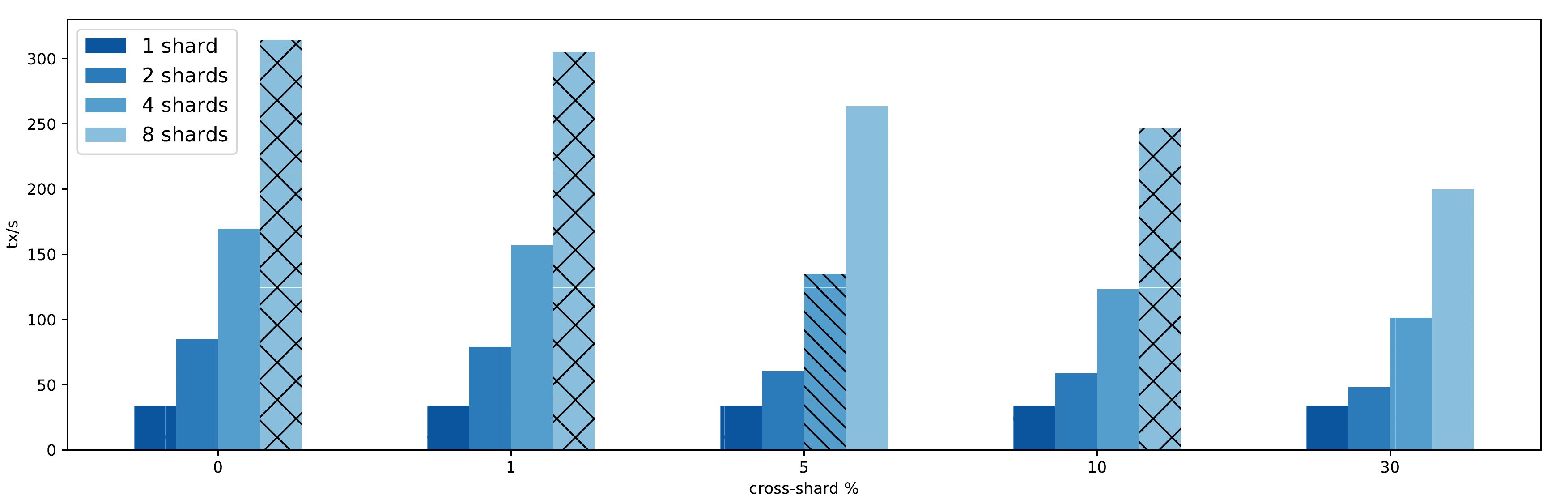}
  \caption{Performance with varying number of shards and different cross-shard transaction rates.}
  \label{fig:aggregated_tput_comparison}
\end{figure*}

Figure~\ref{fig:ck_tput}~(left) shows a nearly linear increase in the average number of
transactions per second as we increase the number of shards, except when there
are eight shards. The reason the throughput with eight shards is lower than
expected is that there were not enough ready-to-run transactions in the dependency
graph, making the client wait for blocked transactions to finish. This is better
visualized in Figure~\ref{fig:ck_tput}~(right), where vertical dashed lines
mark the point when each one of the eight shards had less outgoing transactions
than established at the beginning of the experiment.

To better understand how varying cross-shard transaction rates can affect performance we
conduct experiments in the next section that confirm the performance gains obtained with \cryptokitties.

\subsection{SCoin}\label{sec:scalable_coin_exp}

We now present results for the SCoin application defined in Section~\ref{sec:scalable_coin}.
In these experiments, we benchmark how well the protocol can perform with a
single application with varying number of transactions that require
cross-shard communication (i.e., tokens transferred between different partitions).
We try to measure latency and throughput tradeoffs with a varying number of cross-shard transactions.

Each client in the experiment tries to execute \emph{transfer} transactions in
a closed-loop. If the transaction is cross-shard, i.e., if a client is trying
to transfer its token to an account that resides in a different shard, the
client first move its account to the corresponding shard and then executes the
\emph{transfer} transaction afterward in the destination shard.
Similarly to the previous experiment, we tune the number of clients in the
system to avoid a significant degradation of the average latency for each
client, thus caping each shard to 250 clients.
We experiment with different cross-shard transaction rates for different shard numbers.

Figure~\ref{fig:aggregated_tput_comparison} shows the aggregate throughput of the system for a varying number of shards and different percentages of 
cross-shard transactions.
The one shard experiment is shown in every cross-shard rate experiment as a reference.
We can see how performance degrades when increasing the number of cross-shard transaction rates,
but the throughput grows linearly at different rates of cross-shard.
For comparison, the previous experiment had on average 5.86\%, 7.93\%, 7.85\% cross-blockchain
transaction rate for 2, 4 and 8 shards, respectively.
Any system that reserves part of its allocated resources to
process cross-shard transactions will have similar behavior, due to
cross-shard transactions occupying the resources that otherwise would be used by
single-shard transactions.
%

\begin{figure*}[htb]
  \centering
  \includegraphics[width=0.48\linewidth]{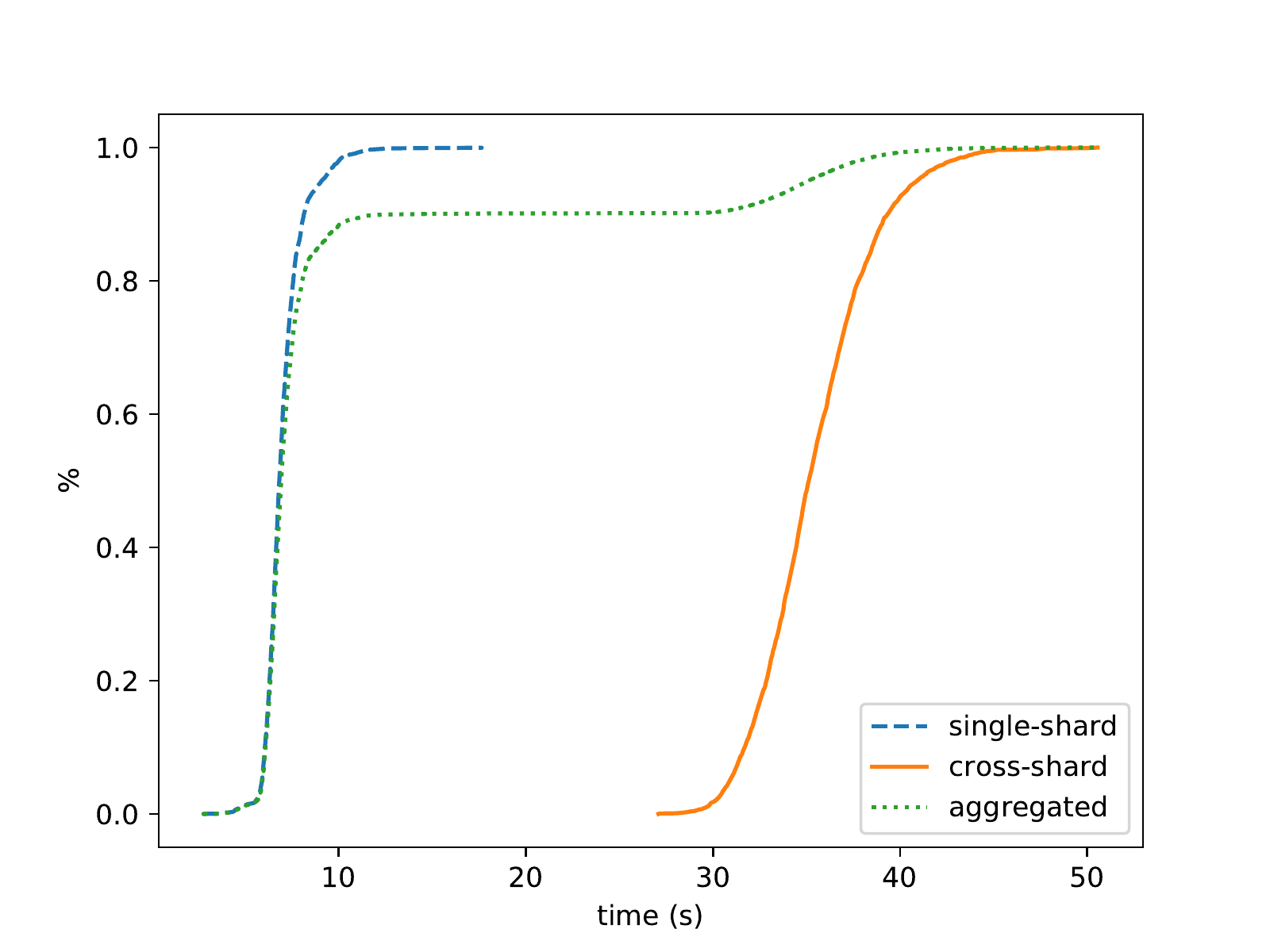}
  \includegraphics[width=0.48\linewidth]{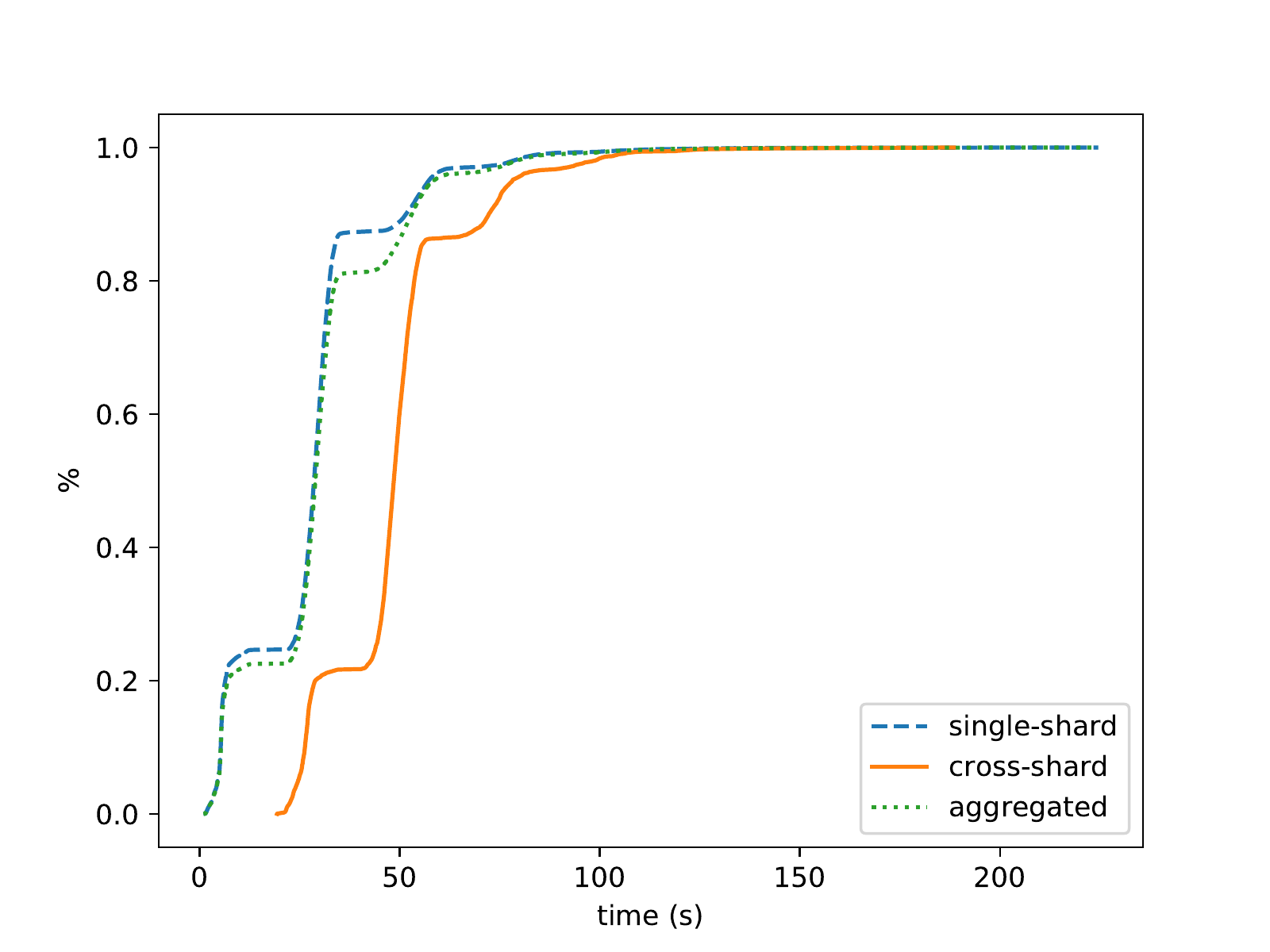}
  \caption{Latency CDF for 4 shards experiment with 10\% cross-shard transactions.}
  \label{fig:cdf_lat_4p_10pc}
  \label{fig:cdf_lat_4p_1pc_retry}
\end{figure*}

The average latency for clients does not change significantly when increasing
the number of shards, remaining at around 7 seconds for single-shard
transactions and 34 seconds for cross-shard transactions.
Cross-shard transactions demand two transactions for each move operation, plus
waiting for two blocks to prove the contract's state and one final transaction
to complete the operation, confirming the expected latency of waiting for five blocks per cross-shard transaction.
In Figure~\ref{fig:cdf_lat_4p_10pc}~(right) we can see the cumulative distribution
function (CDF) for clients' observed latencies in a scenario with four shards and 10\% cross-shard
transactions rate.
%
The aggregated latency shows both single and cross-shard transactions and we can
see that, as expected, around 10\% of the transactions takes more than 30 seconds to complete.
Differently from other sharded systems (e.g.,~\cite{p_store}) the protocol does not suffer from a
\emph{convoy effect}~\cite{convoyeffect}, that is, cross-shard transactions do not delay single-shard transactions.

\subsubsection{Retries}
In the previous experiments, to better control the rate of cross-shard transactions,
clients submit transactions only if they know the contracts are not going to be moved
when the transaction is executed.
To better model transactions in the presence of conflicts we experiment with the \textit{SCoin}
contract without any help from external sources.
Two \replaced{situations can make}{scenarios are possible on which} clients \deleted{have to} retry transactions: when performing a
single-shard transaction and the interacted contract is moved to another shard or
when performing a cross-shard transaction and the called contract is moved to another shard.
In the experiment we make clients wait a random time corresponding to the creation of 0 to 10 blocks
if the transaction fails for any of these reasons, this is done to prevent contracts moving back and forth in an
endless cycle.

\added{Figure~\ref{fig:cdf_lat_4p_1pc_retry}~(right) shows the ideal latency where no conflicts exist, and Figure~\ref{fig:cdf_lat_4p_1pc_retry}~(left) shows the latency when conflicts can happen.}
A clear increase in latency is observed when comparing both figures\added{, but when retries can happen}
\deleted{Figure~\ref{fig:cdf_lat_4p_1pc_retry}~(right) shows a clear increase in latency when compared with the
results showed in Figure~\ref{fig:cdf_lat_4p_10pc}~(left).}
the number of times the same transaction fails and has to retry is highly skewed,
for instance, 66\% of the transactions that retry, do it just once, and
only 1\% of these transactions are retried more than three times.

\section{IBC experiments} \label{sec:ibc_exp}
This section presents some experiments for inter-blockchain communication (IBC) considering Burrow/Tendermint and Ethereum.
These experiments aim to measure the time and gas (which translates to cryptocurrency costs) consumed for operations with five different applications:

\begin{itemize}
  \item \textit{SCoin}: Transfer one token from one blockchain to another and transfer the
  virtual currency to another account in the target blockchain.

  \item \cryptokitties: Transfer one virtual cat from one
  blockchain to another, breeds the cat with an existing cat in the target
  blockchain and gives birth to another cat.

  \item State 1, State 10, State 100: Transfer the state containing respectively 1, 10
  and 100 32-byte state variables from one blockchain to another.
\end{itemize}

\begin{figure*}[htb]
  \centering
  \includegraphics[width=0.48\linewidth]{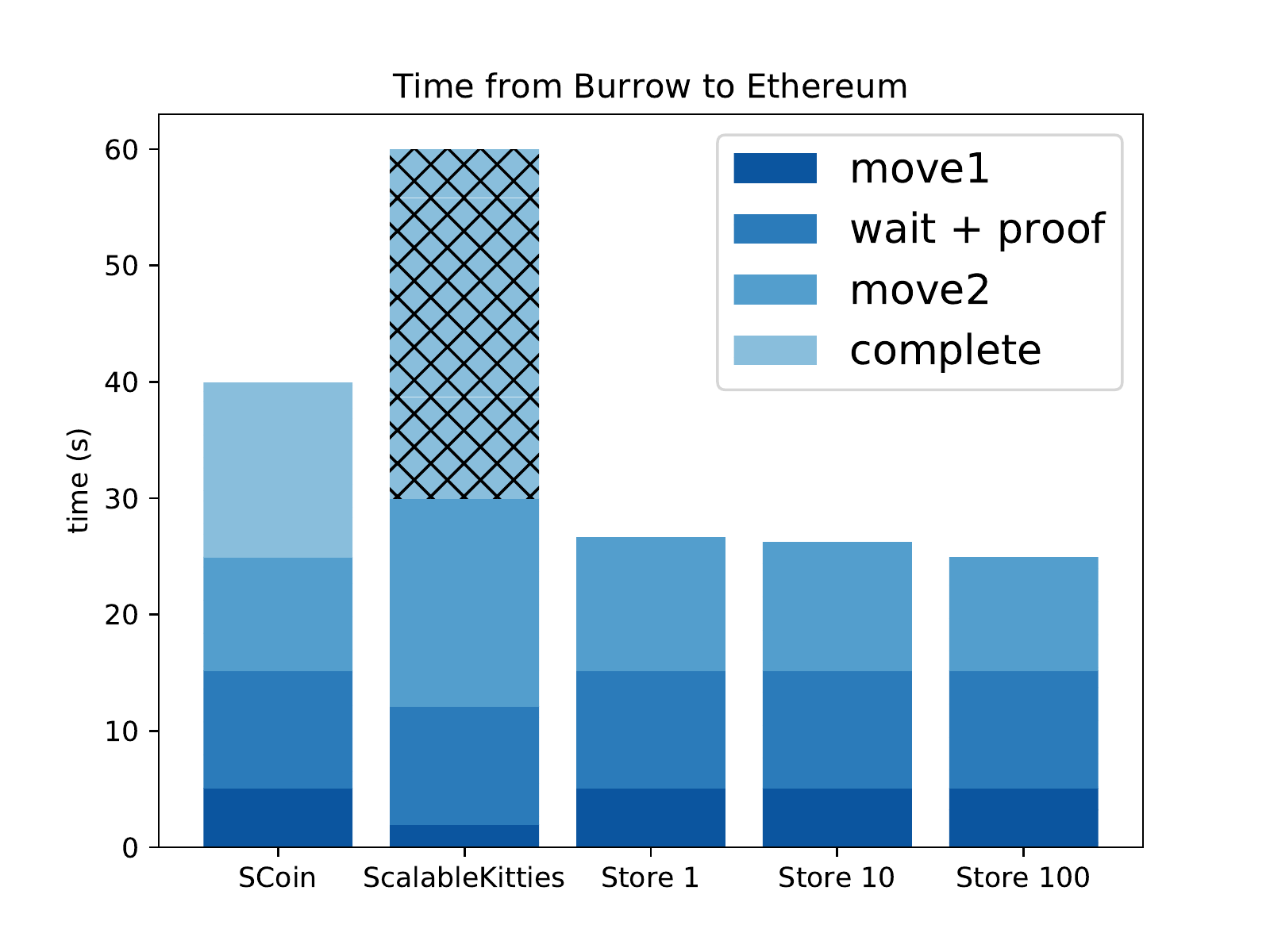}
  \includegraphics[width=0.48\linewidth]{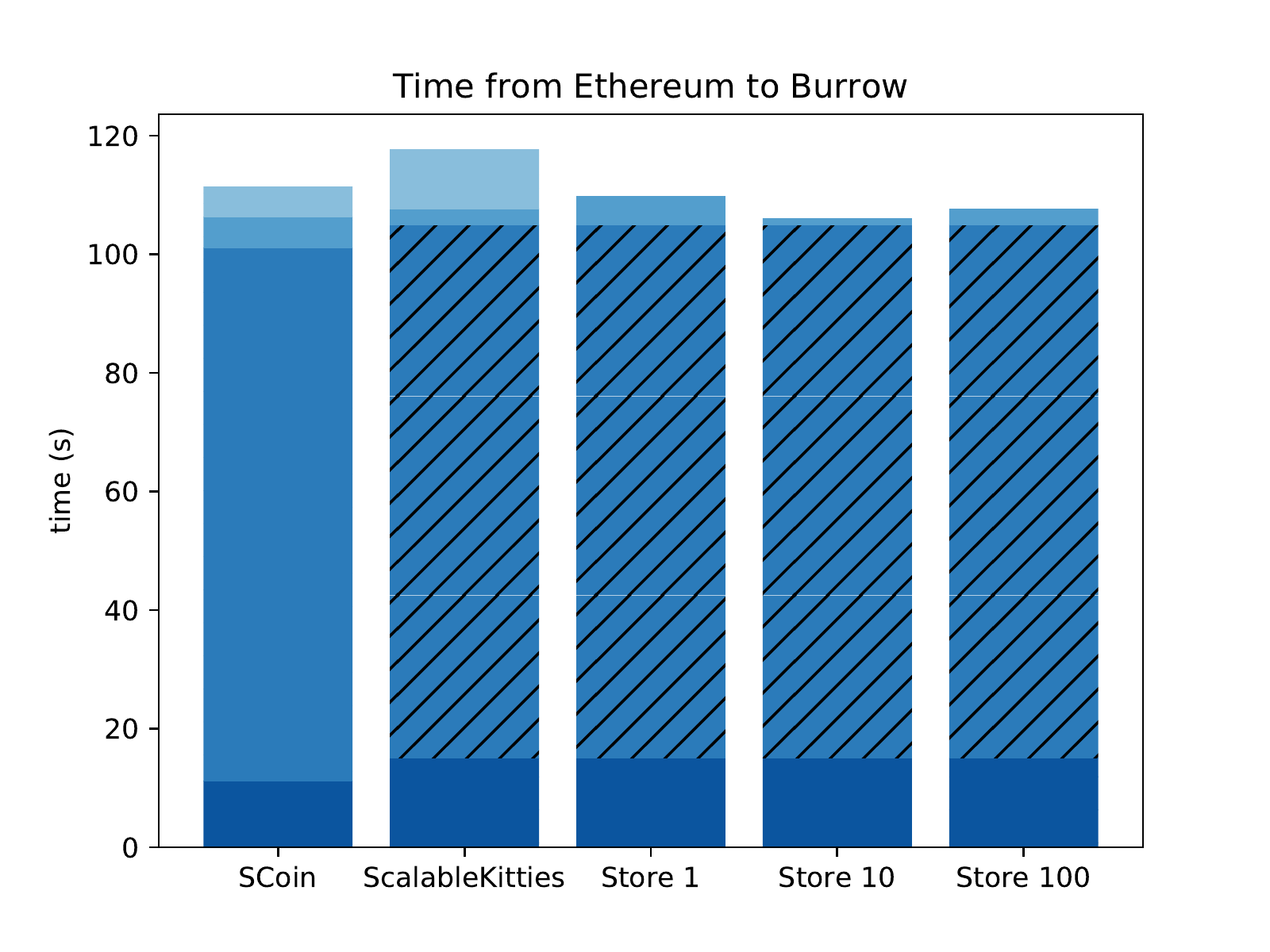}
  \caption{Latency for five different inter-blockchain applications.}
  \label{fig:exp_ibc_time}
\end{figure*}

\begin{figure*}[htb]
  \centering
  \includegraphics[width=0.48\linewidth]{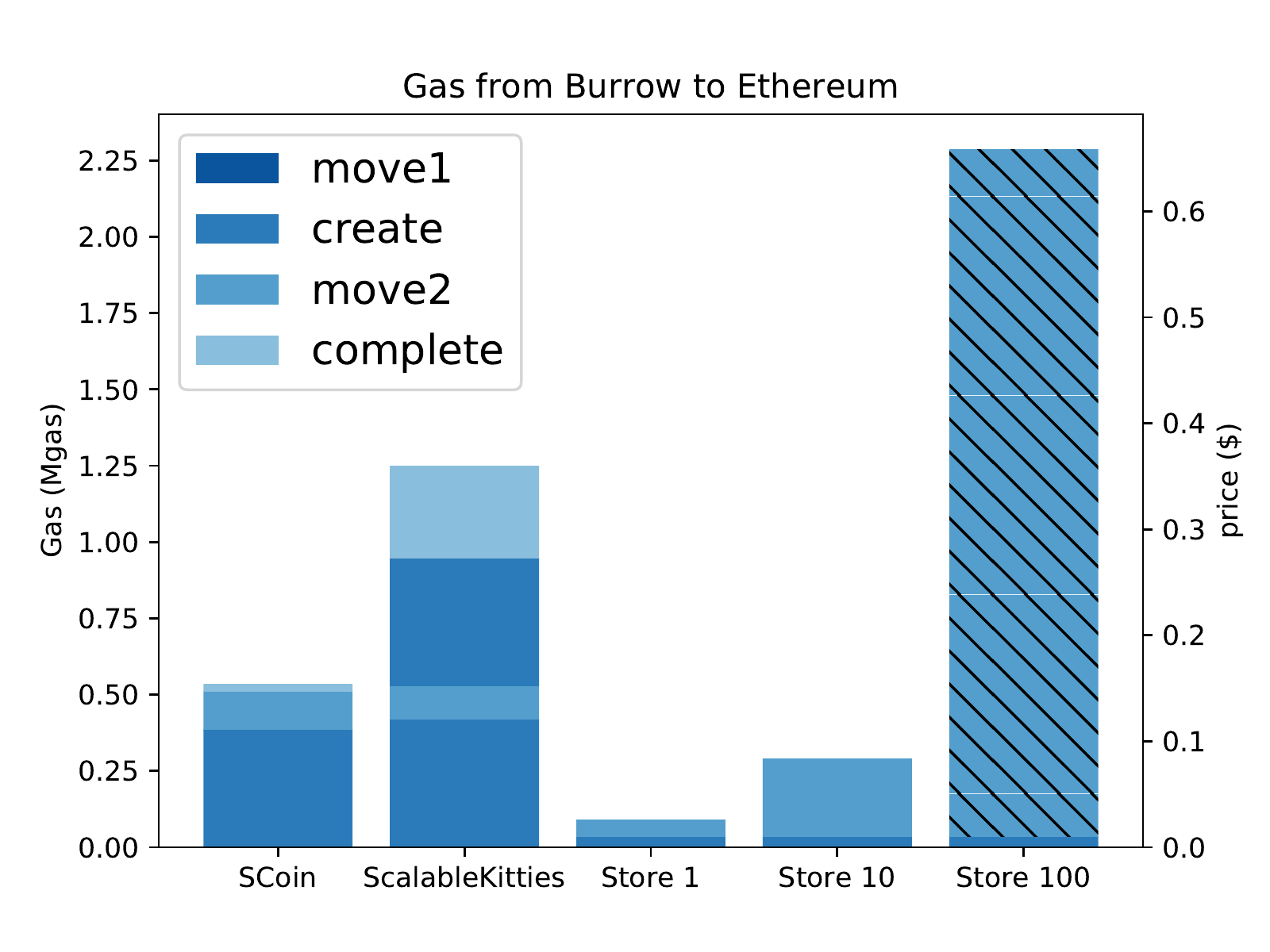}
  \includegraphics[width=0.48\linewidth]{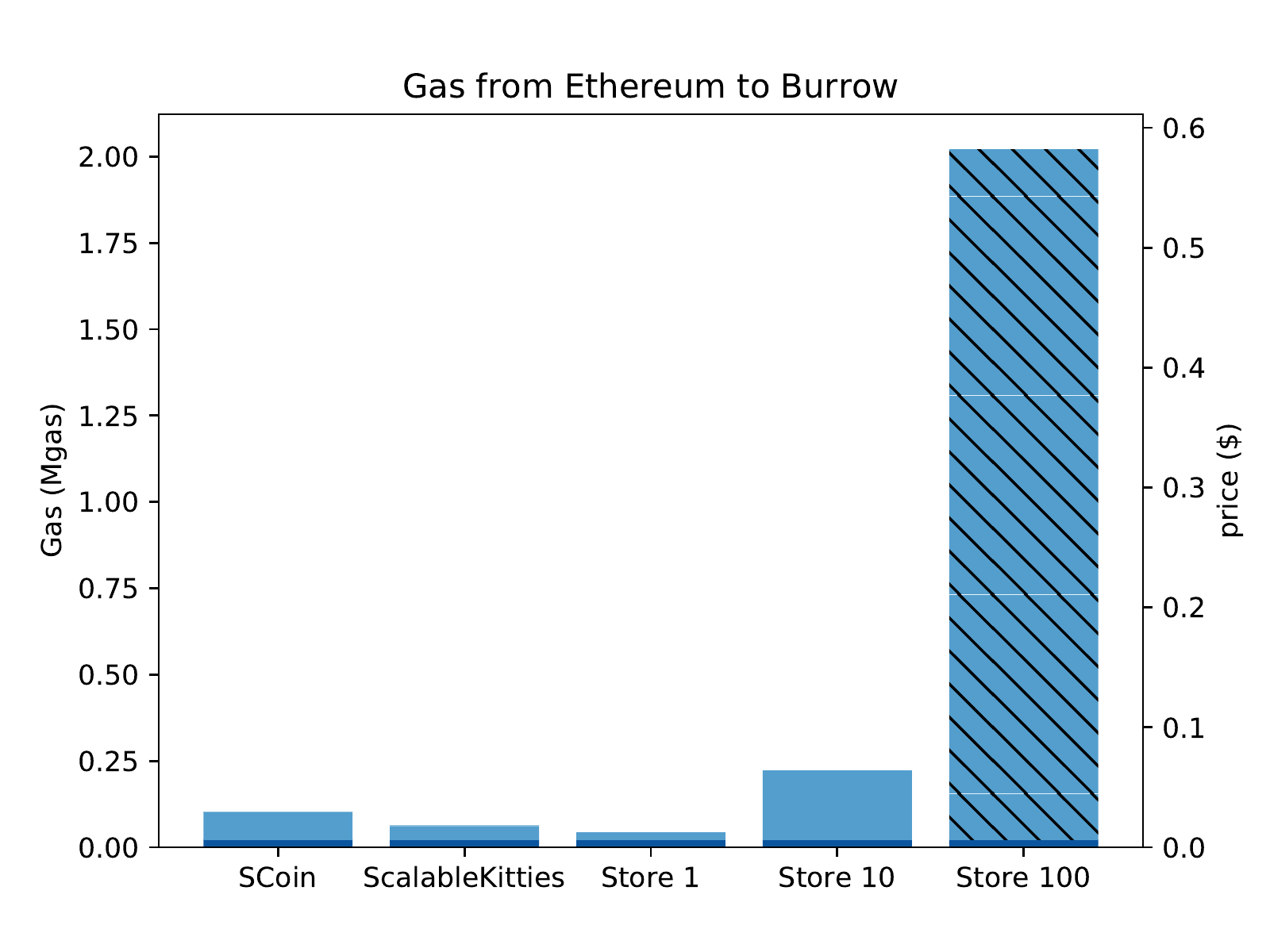}
  \caption{Gas and monetary costs for five different inter-blockchain applications.}
  \label{fig:exp_ibc_gas}
\end{figure*}

These operations require moving their corresponding smart contract from the source to the target blockchain, an operation requiring two transactions
(Move1 and Move2) and the wait for $p$ blocks in-between transactions.
After that, further transactions might need to be executed in the target blockchain.
In our examples: \textit{SCoin} requires one further operation to transfer the token to a contract in the target blockchain, while
\cryptokitties requires two transactions \emph{breed} and \emph{giveBirth} to mate and
produce a new virtual cat, respectively.
The state transfer experiments do not require any further transactions for completion.

Figure~\ref{fig:exp_ibc_time} presents the time required to perform an operation from
Ethereum to Burrow and vice-versa. Unsurprisingly, the time to perform a single transaction
is bound to the latency between consecutive blocks in each blockchain.
To execute Move2 from Ethereum to Burrow one is required to wait for 6 Ethereum blocks that
translates to approximately 90 seconds and ends up dominating the overall time for every
operation.

A good way to analyze the impact of each operation in the system's performance is to measure
the gas consumed by each operation, the gas cost is expected to grow linearly with the size
of the transferred smart contract state, since creating or modifying state variables are expensive 
operations in the EVM model.
Figure~\ref{fig:exp_ibc_gas} shows in the left y-axis the amount of gas paid by each transaction.
To better understand these costs, in the right y-axis we present the same costs in US dollars, considering the current average value of one gas as two Gwei ($2\times 10^{-9}$ Eth) and one Eth as \$\ethprice (the price in the middle of December of 2019).

The way Burrow and Ethereum calculate gas prices are different, besides the actual values per
operation being different, some operations pay no gas, e.g., in Ethereum, if one smart contract
or transaction creates a new smart contract it pays an amount of gas per byte of the contract code that is being created,
while in Burrow no gas is paid per byte of code.
In both systems the code is immutable and stored in the state Merkle-tree with the key based on the code's hash.

In Figure~\ref{fig:exp_ibc_gas}, the vertical hatched bars represent the gas paid
by the creation of new contracts in Ethereum, every recreated contract pays a constant gas based on the
size of the moved code. In the \cryptokitties application the \emph{giveBirth} function creates
a new contract thus it pays for the gas again, for \textit{SCoin} and \cryptokitties the gas paid for the code
creation corresponds to around 70\% of the total gas cost.
We note that it is possible to reduce significantly the Ethereum contract creation costs if the contract code is already in the blockchain.

\section{Related work} \label{sec:related_work}

Scaling blockchains is a hot topic for both industry and academia, \replaced{and}{with} many proposals \replaced{have appeared in the last years (e.g., \cite{possidechains},
\cite{sok}).}{with many proposals appearing in the last years.}
\replaced{In this paper, we showed that scalability can be achieved with a novel IBC protocol and sharding.}{In this paper we showed such scalability can be achieved with a novel IBC protocol and sharding.}
A different approach \replaced{to}{for} scaling blockchains is to shift computation from the blockchain (on-chain)
to the outside (off-chain). For example, in the method proposed
by TrueBit~\cite{truebit} most of the computation is done off-chain\replaced{.
Cheating}{, cheating}
participants can be caught by using an on-chain game in which anyone can
prove they misbehaved in logarithmic time.
Another example of a scaling solution is the lightning network~\cite{lightning},
which works on top of Bitcoin, and its Ethereum \replaced{counterpart}{counterparty},
Raiden~\cite{raiden}.
Users of such systems create off-chain channels between them in order to minimize on-chain transactions.
The efficiency of such systems is highly application-dependent.

Pegged Sidechains~\cite{pegged_sidechains} focus on transferring assets between proof-of-work blockchains. The idea is to lock assets in one blockchain and recreate them in another blockchain by providing a proof of such locking in the original blockchain. The Move protocol generalizes Pegged Sidechains in several aspects. First, it allows to transfer any state across blockchain systems, not only assets; second, it applies to both proof-of-stake and proof-of-work approaches; third, the Move protocol shifts control to the application developer, who can develop scalable applications with their own logic (e.g., introducing load balancing mechanisms).

In HyperService~\cite{hyperservice}, the authors create a new programming language and system to make blockchain applications interoperable. HyperService orchestrates the execution of applications that span multiple blockchains. We take a different approach: a smart contract is executed in one blockchain, after the smart contract dependencies are moved to the same blockchain.

In \deleted{``Atomic cross-chain swaps''~} \cite{atomic_cross_shard}, atomic token swaps
between multiple blockchains are studied and proposed as a protocol that,
similar to ours, is done in two phases.
As noted by the author, ``atomic
cross-chain swap is an atomic cross-chain transaction, but not vice-versa''.
Our protocol could be used to implement atomic swaps in a similar way as shown in
\ref{sec:atomic_swap}, although a more efficient solution for performing
token swaps with more than two blockchains, combining our protocol with the techniques 
proposed \replaced{in \cite{atomic_cross_shard}}{by the author,} would be interesting future work.
In \cite{abebe2019enabling} the authors propose a solution for permissioned ledgers
where blockchains implement a common ``relay'' mechanism for cross-blockchain transactions
with smart contracts. The protocol provides a great deal of flexibility but
it requires smart contract developers to have a deep understanding of the underlying cross-chain protocol.
Without allowing contract's state to migrate, blockchain systems risk having their
performance limited by cross-blockchain transaction performance.

PolkaDot~\cite{polkadot} aims to create a decentralized \emph{federation} of blockchains
by allowing other blockchains (called para-chains) to exist. Existing blockchains can
be interfaced by \emph{relay-chains}, but no details are given on how the validation happens on
existing blockchains. Similar to our work, blockchains in PolkaDot need well-defined 
parameters so they can interoperate, e.g., the number of blocks to wait to accept
the transaction as being final.
Cosmos~\cite{cosmos} also aims to provide IBC by allowing multiple blockchains,
called \emph{zones}, to communicate with each other. All Cosmos zones run the
Tendermint algorithm for consensus. One zone, called \emph{Cosmos hub} acts as
a central communication interface for all other zones. \deleted{It is not clear how
existing blockchains can participate in the Cosmos protocol.}
%
Interledger~\cite{interledger} is a proposed protocol for IBC that tackles
payments.
To achieve safety and liveness, transactions are either escrowed by notaries
that run PBFT~\cite{pbft} or use incentives on rational actors \deleted{but requires
synchronous communication}. The Interledger approach does not integrate with existing blockchains
and is constrained to simple token transfers.

In Omniledger~\cite{omniledger}, the authors developed a protocol that can
process transactions across shards, called Atomix, built on top of
Bitcoin's UTXO model. In Atomix, clients can lock their input and are left
with the burden of unlocking them in case the transaction aborts.
The authors briefly discuss applying Atomix to smart
contracts and suggest \added{that} Atomix is suitable for scenarios where clients execute
simple operations and transactions do not conflict.
Our approach exposes IBC primitives to developers to give them
more freedom to programmatically condition cross-blockchain operations.
Similar to Omniledger, Elastico~\cite{elastico} focus on Bitcoin's UTXO model
and there is no need for cross-shard transactions because outputs are assumed to be
disjointedly distributed to shards.

Chainspace~\cite{chainspace} builds on top of BFT-SMaRt~\cite{bftsmart}, an
open source BFT consensus implementation written in Java. 
One of the subsystems of Chainspace is S-BAC, a two-phase commit protocol that deals with cross-shard transactions.
In our model, we expose sharding primitives that let developers
programmatically control the shard designation of smart contracts, doing so
simplifies the protocol and allow for a more organic distribution of objects.
Differently from S-BAC, our protocol does not implement aborts and it is the
developer's responsibility to avoid contracts stuck on the first phase of the
protocol.
Chainspace provides helpful insights for further works on sharding smart
contracts that could be also applied in the presented protocol, e.g., having
shards checkpoints can be beneficial to reduce the bandwidth and storage cost
of moving a contract. \added{Similarly, the efficient shard state transfer protocol described in \cite{elastic_smr} can alleviate this costs}

A protocol \added{with similar goals as ours} has been proposed in the Ethereum research forum
concurrently with the development of this work~\cite{yanking}. It describes a similar
operation to move the state from one shard to another, called \emph{yanking}. Other
threads in the same forum further extend the proposed idea, but until the
writing of this paper, it remains a work in progress tailored specifically
for the Ethereum environment, \added{and} constrained by its own design choices.
\deleted{We believe our work sheds some light on the matter by providing a more
generic approach to this problem and evaluate its costs.}

\section{Conclusions} \label{sec:conclusions}
In this paper, we present a practical protocol that can be used to develop smart contracts that can move between different blockchains.
Our protocol enables smart contract developers to create blockchain applications that interoperate and scale\deleted{-out} in an ecosystem of multiple blockchains.
We developed two applications for our protocol and extensively evaluated their performance and tradeoffs.
The simplicity of our protocol opens up possibilities for further improvements, such as decentralized load balancing smart contracts for sharded blockchains.

\section*{Acknowledgments}

We wish to thank our shepherd, Pascal Felber, and the anonymous reviewers for helping us improve the paper.
This work was partially supported by FCT through project ThreatAdapt (FCT-FNR/0002/2018), the LASIGE Research Unit (UIDB/00408/2020 and UIDP/00408/2020), and the Swiss National Science Foundation (project number 175717).

\bibliographystyle{IEEEtran}
\bibliography{references}

\end{document}